\documentclass{rsproca}

\usepackage{natbib}
\usepackage{graphicx}
\usepackage{amsmath}
\usepackage{rotating}
\usepackage{afterpage}
\usepackage{pgfplots}
\usepackage{xargs}
\usepackage[multidot]{grffile}

\newcommand{\plotgrid}[9]{%
    \def\tempa{#1}
    \def\tempb{#2}
    \def\tempc{#3}
    \def\tempd{#4}
    \def\tempe{#5}
    \def\tempf{#6}
    \def\tempg{#7}
    \def\temph{#8}
    \def\tempi{#9}
    \plotgridcontinued
}
\newcommandx{\plotgridcontinued}[5]{
\begin{figure}
\centering
\begin{tikzpicture}
    \begin{axis}[
        axis x line=bottom,
        axis y line=left,
        width=\textwidth,
        height=1.25*\textwidth,
        axis equal=false,
        enlargelimits=false,
        xmin=0.5,
        xmax=3.5,
        ymin=0.5,
        ymax=4.5,
        xlabel={Initial embryo mass, M$_e$ (M$_\oplus$)},
        ylabel={Ratio of the total mass in the embryos to the total mass in the planetesimals, $\Sigma M_e$:$\Sigma M_p$},
        ytick={1,2,3,4},yticklabels={$1$:$1$,$2$:$1$,$4$:$1$,$8$:$1$},
        xtick={1,2,3},xticklabels={$0.025$,$0.05$,$0.08$}
        ]
      \addplot graphics[xmin=0.55,xmax=1.45,ymin=3.55,ymax=4.45]{\tempa};
      \addplot graphics[xmin=1.55,xmax=2.45,ymin=3.55,ymax=4.45]{\tempb};
      \addplot graphics[xmin=2.55,xmax=3.45,ymin=3.55,ymax=4.45]{\tempc};
      \addplot graphics[xmin=0.55,xmax=1.45,ymin=2.55,ymax=3.45]{\tempd};
      \addplot graphics[xmin=1.55,xmax=2.45,ymin=2.55,ymax=3.45]{\tempe};
      \addplot graphics[xmin=2.55,xmax=3.45,ymin=2.55,ymax=3.45]{\tempf};
      \addplot graphics[xmin=0.55,xmax=1.45,ymin=1.55,ymax=2.45]{\tempg};
      \addplot graphics[xmin=1.55,xmax=2.45,ymin=1.55,ymax=2.45]{\temph};
      \addplot graphics[xmin=2.55,xmax=3.45,ymin=1.55,ymax=2.45]{\tempi}; 
      \addplot graphics[xmin=0.55,xmax=1.45,ymin=0.55,ymax=1.45]{#1};
      \addplot graphics[xmin=1.55,xmax=2.45,ymin=0.55,ymax=1.45]{#2};
      \addplot graphics[xmin=2.55,xmax=3.45,ymin=0.55,ymax=1.45]{#3};                  
      \end{axis}
      \end{tikzpicture}
      \caption{#4}
      \label{#5}
      \end{figure}
 }

\begin{document}

\title{Lunar and Terrestrial Planet Formation in the Grand Tack Scenario}

\author{
S. A. Jacobson$^{1,2}$ and A. Morbidelli$^{1}$}

\address{$^{1}$Observatoire de la C{\^o}te d'Azur, Laboratoire Lagrange, Bd. de l'Observatoire, B.P. 4029, F-06304 Nice Cedex 4, France\\
$^{2}$Universit{\"a}t Bayreuth, Bayerisches Geoinstitut, D-95440 Bayreuth, Germany }

\subject{xxxxx, xxxxx, xxxx}


\corres{Seth A. Jacobson\\
\email{seth.jacobson@oca.eu}}

\begin{abstract}
We present conclusions from a large number of N-body simulations of the giant impact phase of terrestrial planet formation. We focus on new results obtained from the recently proposed Grand Tack model, which couples the gas-driven migration of giant planets to the accretion of the terrestrial planets. The giant impact phase follows the oligarchic growth phase, which builds a bi-modal mass distribution within the disc of embryos and planetesimals. By varying the ratio of the total mass in the embryo population to the total mass in the planetesimal population and the mass of the individual embryos, we explore how different disc conditions control the final planets. The total mass ratio of embryos to planetesimals controls the timing of the last giant (Moon forming) impact and its violence. The initial embryo mass sets the size of the lunar impactor and the growth rate of Mars. After comparing our simulated outcomes with the actual orbits of the terrestrial planets (angular momentum deficit, mass concentration) and taking into account independent geochemical constraints on the mass accreted by the Earth after the Moon forming event and on the timescale for the growth of Mars, we conclude that the protoplanetary disc at the beginning of the giant impact phase must have had most of its mass in Mars-sized embryos and only a small fraction of the total disc mass in the planetesimal population. From this, we infer that the Moon forming event occurred between $\sim$60 and $\sim$130~My after the formation of the first solids, and was caused most likely by an object with a mass similar to that of Mars.
\end{abstract}


\begin{fmtext}

\end{fmtext}


\maketitle

\section{Introduction}
Terrestrial planet formation has been investigated over several decades from the first semi-analytic approaches~\citep{Safronov:1969cg} to the most sophisticated numerical simulations~\citep[e.g.][to quote just a few]{Kokubo:2010im,Morishima:2010cs,Chambers:2013cp}. The basic ideas, sequence of events leading to the formation of the terrestrial planets were laid down by G. Wetherill~\citep[see][for early reviews]{Wetherill:1980iq,Wetherill:1990fe}. Wetherill was also a pioneer in understanding that the evolution of the asteroid belt and the accretion of the terrestrial planets are intimately related~\citep{Wetherill:1992ja}. The possibility for the delivery of water to the Earth by primitive asteroids was later explored in~\citet{Morbidelli:2000ex} and in subsequently by others~\citep[e.g.][]{Raymond:2004de,Raymond:2007ks,OBrien:2014vp}. 

In Wetherill's paradigm, the solids, initially dust particles, coagulate to form planetesimals. The collisional evolution of these planetesimals in a disc that starts dynamically cold (i.e. with very small orbital eccentricities and inclinations) leads to the formation of a limited number of more massive bodies called ``planetary embryos,'' by the processes known as {\it runaway growth}~\citep{Greenberg:1978ea,Wetherill:1989ev} and {\it oligarchic growth}~\citep{Kokubo:1998ka}. The embryos then collide with each other, leading to the assemblage of the most massive terrestrial planets (Earth and Venus) on longer timescales through giant impacts like that postulated for the origin of the Moon~\citep{Hartmann:1975je}. Numerical simulations have allowed a progressively more detailed quantitative investigation of this process. The first modern numerical simulations were performed by~\citet{Chambers:1998em}, followed by~\citet{Agnor:1999ha},~\citet{Chambers:2001kt},~\citet{Chambers:2002hj},~\citet{Levison:2003fc},~\citet{Raymond:2004de} and many others~\citep[see][for a review of the evolution of the field]{Morbidelli:2012iz}. 

Overall the simulations have shown that Wetherill's paradigm is successful. A handful of planets are formed in the terrestrial planet region and only rarely in the asteroid belt~\citep{Raymond:2009is}. The largest planets have masses comparable to Earth and Venus. The early simulations produced planets systematically with orbits too  eccentric and inclined relative to the real ones, but this problem eventually disappeared due to the inclusion in the simulations of a substantial amount of planetesimals, producing dynamical friction (i.e. the damping of eccentricities and inclinations through gravitational interactions) on planetary embryos and final planets~\citep{OBrien:2006jx}. The timescale of terrestrial planet accretion is tens of millions of years, broadly consistent with that suggested by isotopic chronometers~\citep[see][for a review]{Kleine:2009tp}. Giant impacts are frequent, so that the Moon forming event is not an oddity.~\citet{Agnor:1999ha} did a detailed analysis of the giant impacts occurring in the simulations. They showed that impacts qualifying for a Moon forming event, defined as those carrying an angular momentum relative to the Earth comparable to the current angular momentum of the Earth-Moon system, are not rare. Such impacts typically involve projectiles of 1--2 Mars masses and occur tens of millions of years after the beginning of the simulation, which is in quantitative agreement with the characteristics inferred for the Moon forming collision~\citep{Canup:2001te}. 

Nevertheless, the simulations of terrestrial planet formation starting from a disc of embryos and planetesimals extending from the Sun to the orbit of Jupiter have some drawbacks. The most important one is that the synthetic planet produced in the simulations at $\sim$1.5 AU is systematically much more massive than Mars~\citep{Chambers:2001kt}.~\citet{Raymond:2009is} did a detailed analysis of this issue looking at the dependence of the results on the initial orbit of the giant planets. They found that only simulations with Jupiter and Saturn on orbits with the current semi-major axes but significantly larger eccentricities can form a small Mars. This is due to the presence of a strong secular resonance in the vicinity of 1.5 AU, which depletes most of the material from the feeding zone of the planet. The location of this resonance is very sensitive to the orbital separation between Jupiter and Saturn, so the assumption that the planets occupy their current orbits is important. However, as also discussed in~\citet{Raymond:2009is}, this initial condition for the orbits of the giant planets seems unrealistic. In fact, no mechanism has ever been found to excite the orbits of Jupiter and Saturn to the required values and leave the giant planets at current orbital separations. The so-called {\it Nice} model provides an excitation mechanism, but said excitation is expected to have occurred much later than terrestrial planet formation in order to explain the Late Heavy Bombardment of the Moon~\citep{Gomes:2005gi} and the formation of the Oort cloud in a dispersed galactic environment~\citep{Brasser:2013dw}. Moreover, even accepting for sake of argument that the excitation event occurred as soon as the gas was removed from the disc, the magnitude of the excitation would not have significantly exceeded the current eccentricity of Jupiter~\citep{Nesvorny:2012fy}. Finally, even accepting that Jupiter and Saturn had acquired more eccentric orbits by some unknown mechanism, the damping of their orbital eccentricities to the current values due to planetesimal scattering would have forced some radial migration of the planets to more separated orbits. Thus assuming {\it larger} eccentricities and {\it current} semi-major axes is inconsistent. Clearly, this is a dead end. 

A novel idea was proposed by~\citet{Hansen:2009ke}. He showed that the key parameter for obtaining a small Mars is the radial distribution of the solid material in the disc. If the outer edge of the disc of embryos and planetesimals is at about 1 AU, with no solid material outside of this distance, even simulations with giant planets on circular orbits systematically produce a small Mars (together with a big Earth). The issue is then how to justify the existence of such an outer edge and how to explain its compatibility with the existence of the asteroid belt between 2 and 4 AU. The asteroid belt has today a very little total mass~\citep[about $6\times 10^{-4}$ Earth masses;][]{Krasinsky:2002ku}, but it must have contained at least a thousand times more solid material when the asteroids formed~\citep{Wetherill:1989ev}. Nevertheless, the spectacular success of Hansen's simulations motivated research to understand a plausible explanation of his initial conditions, coherent with the observed structure of the asteroid belt and the outer Solar System. This led to the definition of the so-called {\it Grand Tack} scenario~\citep{Walsh:2011co}. 

In the next section, we briefly review this scenario and the results published so far. Then, in sections 3-6, we present new results devoted to characterizing the initial disc of embryos and planetesimals and to constrain the timing of lunar formation as well as the most likely mass and velocity of the projectile. 

\section{The Grand Tack scenario}
The Grand Tack model is a giant planet migration and terrestrial planet formation paradigm designed to reproduce the structure of the inner Solar System, particularly the mass of Mars~\citep{Walsh:2011co}. It is the first model where the giant planets are not assumed to be on static orbits. Instead~\citet{Walsh:2011co} studied the co-evolution of the orbits of the giant planets and of the planetesimal and embryo precursors of the terrestrial planets, during the era of the disc of gas.~\citet{Walsh:2011co} built their model on previous hydro-dynamical simulations showing that the migration of Jupiter can be in two regimes: when Jupiter is the only giant planet in the disc, it migrates inwards~\citep{Lin:1986ef}, but when paired with Saturn both planets typically migrate outward, locked in a 2:3 mean motion resonance~\citep[where the orbital period of Saturn is 3/2 of that of Jupiter;][]{Masset:2001dk,Morbidelli:2007ew}. Thus, assuming that Saturn formed later than Jupiter,~\citet{Walsh:2011co} envisioned the following scenario: first, Jupiter migrated inwards while Saturn was still growing; then, when Saturn reached a mass close to its current one, it started to migrate inwards more rapidly than Jupiter, until it captured the latter in the 3/2 resonance~\citep{Masset:2001dk,Pierens:2008jg}; finally, the two planets migrated outwards together until the complete disappearance of the gas in the disc. The final orbits of the giant planets are consistent with the initial conditions of the most recent version of the Nice model~\citep{Morbidelli:2007dc,Levison:2011gt,Nesvorny:2012fy} that explains the final transition towards the current giant planet orbits, hundreds of My after the disappearance of the disc of gas. 

In the Grand Tack scenario, the extent of the inward and outward phases of migration of Jupiter are unconstrained a priori, because they depend on properties of the disc and of giant planet accretion that are unknown, such as the time-lag between Jupiter's and Saturn's formation, the speed of inward migration (which depends on the disc's viscosity), the speed of outward migration (which depends on the disc's scale height), and the time-lag between the capture in resonance of Jupiter and Saturn and the photo-evaporation of the gas. However, the extent of the inward and outward migration of Jupiter can be deduced by looking at the resulting structure of the inner Solar System. In particular,~\citet{Walsh:2011co} showed that a reversal of Jupiter's migration at 1.5 AU would provide a natural explanation for the existence of the outer edge at 1 AU of the inner disc of embryos and planetesimals, required to produce a small Mars as in Hansen's model. Because of the prominent reversal of Jupiter's migration that it assumes, the~\citet{Walsh:2011co} scenario is nicknamed the {\it Grand Tack}.

A crucial diagnostic of the Grand Tack scenario is the survival of the asteroid belt. Given that Jupiter should have migrated through the asteroid belt region twice, first inwards, then outwards, one could expect that the asteroid belt should now be totally empty.  However, the numerical simulations by~\citet{Walsh:2011co} show that the asteroid belt is first fully depleted by the passage of the giant planets, but then, while Jupiter leaves the region for the last time, it is re-populated by a small fraction of the planetesimals scattered by the giant planets during their migration. In particular, the inner asteroid belt is repopulated mainly by planetesimals that were originally inside the orbit on which Jupiter formed, while the outer part of the asteroid belt is repopulated mainly by planetesimals originally in between and beyond the orbits of the giant planets. Assuming that Jupiter initially formed at the location of the snow line, it is then tempting to identify the planetesimals originally closer to the Sun with the anhydrous asteroids of E- and S-type and those originally in between and beyond the orbits of the giant planets with the primitive C-type asteroids. With this assumption, the Grand Tack scenario can explain the co-existence of asteroids of so dramatically different physical properties in the main belt. Thus, following the footsteps of~\citet{Wetherill:1992ja}, the Grand Tack scenario successfully explains both the terrestrial planets and the asteroid belt in a unitary framework. 

We think that the asteroid belt structure is a strong argument in support of the Grand Tack scenario. As an alternative, one could imagine, for instance, that the outer edge at 1 AU of the planetesimal disc was simply the consequence of the inward migration of planetesimals and embryos due to gas-drag and disc tides. A similar concentration of mass towards the inner part of the disc has been invoked to explain the in-situ accretion of the systems of hot Earths and super-Earths recently discovered around other stars~\citep{Hansen:2013fq}. So, why not having a similar process in the Solar System? We think that the asteroid belt rules out this possibility. The inward migration of small planetesimals and large embryos could explain the mass deficit of the asteroid belt, but not its orbital distribution. In absence of the Grand Tack migration of Jupiter,~\citet{OBrien:2007ex} demonstrated that the only mechanism that could give to the belt an orbital structure similar to the observed one is that of mutual scattering of resident embryos~\citep{Wetherill:1992ja,Petit:2002vm}. But if this was the case, then the mass distribution could not be concentrated within 1 AU. Thus, at the current state of knowledge, only the Grand Tack scenario seems able to explain the required mass concentration. 

We also notice that strong claims made at conferences for alternative explanations for the small mass of Mars, not requiring the truncation of the disc~\citep[e.g.][]{Minton:2011tn} have been subsequently dialed back after having investigated more deeply the underlying processes~\citep{Minton:2014du}. 

The Grand Tack simulations of~\citet{Walsh:2011co} reproduce the mass-orbit distribution of the real terrestrial planets of the Solar System like in~\citet{Hansen:2013fq}. The orbital properties of the planets, in terms of orbital excitation and mass concentration are also reproduced quite satisfactorily~\citep[see][and Figure~\ref{fig:scsd}]{Hansen:2013fq}. The scattering towards the inner Solar System of bodies originally located beyond Saturn during the phase of outward migration of the planets also explains the delivery of water to the terrestrial planets~\citep{OBrien:2014vp} in a quantity consistent with geochemical constraints~\citep{Marty:2012gi}.

Concerning the Moon forming event, the simulations presented in~\citet{Hansen:2009ke},~\citet{Walsh:2011co} and~\citet{OBrien:2014vp} show the rapid formation of the Earth. The median timescale for the accretion of 90\% of the Earth's mass is less than 20 My in~\citet{Hansen:2009ke}. Most of the giant impacts in the simulations of~\citet{Walsh:2011co} and~\citet{OBrien:2014vp} occur earlier than 30 My; only a few synthetic Earths experience a last giant impact after 50 My from the beginning of the simulation. This result is intriguing. The timing of the Moon forming event is vastly debated in the geochemical community. Model ages based on radioactive chronometers suggest different times, from 30 My~\citep{Yin:2002ex,Jacobsen:2005bj,Taylor:2009jr} to 60--100 My~\citep{Tera:1973tf,Carlson:1988jz,Touboul:2007is,Halliday:2008bo,Allegre:2008iq,Jacobson:2014cm}. Thus, the Grand Tack model, in its first published  results, seems to argue in favor of the earliest date, or even before. An indication that something may not be right in the published Grand Tack results has been pointed out in~\citet{OBrien:2014vp}: in most of the simulation, characterized by a last giant impact at a time of 30 My or earlier, the Earth subsequently accretes quite a significant amount of mass from planetesimals. This seems at odds with the amount of highly siderophile elements in the Earth's mantle, which suggest that less than 1\% of the Earth's mass was accreted from chondritic material since the Moon forming event. 

For these reasons, we have decided to explore further the potential of the Grand Tack scenario, focusing on the properties of the initial disc of embryos and planetesimals. Some of the conclusions are presented in~\citet{Jacobson:2014cm}. We repeat them below, while reporting also a wealth of new complementary results that could not find space in the format of that paper. The new results constrain the properties of the disc and in turn provide strong indications on the timing of the Moon forming event and the characteristics of the collision. 

\section{New Grand Tack simulation suites}
The new simulations that we have performed follow strictly the recipe of~\citet{Walsh:2011co} in terms of radial extent of the disc, migration of the giant planets and prescription for gas drag. But we have decided to explore the properties of the disc of planetesimals and embryos. In particular we have varied the ratio between the total mass in embryos and in planetesimals and the mass of the individual embryos. 

The rationale for changing these two parameters is that their values reflect the maturity of the process of oligarchic growth. More precisely, the mass of the individual embryos is reflective of the age of the disc. In an older disc the average oligarch will have progressed further towards the isolation mass than in a younger disc, resulting in larger initial embryo masses within the Grand Tack paradigm. Since the earlier evolution from pebbles to planetesimals to embryos is not well understood, as we explore initial conditions for the Grand Tack paradigm we are simultaneously exploring constraints on these earlier epochs. 

Similarly, the ratio of the sum of the mass in embryos to the sum of the mass in planetesimals $\Sigma M_e$:$\Sigma M_p$ is a product of the amount of evolution within the planetesimal disc during the oligarchic growth epoch. During runaway growth, the total mass in embryos quickly grows relative to the total mass in planetesimals. As oligarchic growth proceeds, the total mass in embryos slowly grows due to planetesimal accretion while the total mass in planetesimals declines due to both accretion onto embryos and, perhaps more importantly, collisional grinding within the planetesimal population~\citep{Levison:2012wp}. Therefore, the ratio of the sum of embryo mass to planetesimal mass $\Sigma M_e$:$\Sigma M_p$ during oligarchic growth starts low and increases with time during this phase. Again, as we explore this initial condition for the Grand Tack paradigm, we are simultaneously exploring constraints on evolution within the oligarchic growth regime. 

We explore a large portion of parameter space by adjusting these two initial conditions.~\citet{Walsh:2011co} and~\citet{OBrien:2014vp} focussed on scenarios that begin with $0.025$ and $0.05$ M$_\oplus$ embryos and $\Sigma M_e$:$\Sigma M_p$ of 1:1. We extend this grid of values to include $0.08$ M$_\oplus$ embryos and then we explore a variety of $\Sigma M_e$:$\Sigma M_p$ including 2:1, 4:1 and 8:1. Each suite of simulations is composed of $\sim10$ runs with those initial conditions, with the exception of the 1:1-0.025 and 1:1-0.05 suites, which are taken directly from~\citet{Walsh:2011co} and~\citet{OBrien:2014vp} (4 from each so 8 total per suite). We refer to each suite of simulations as $\Sigma M_e$:$\Sigma M_p$ - $M_e$ so the 1:1-0.025 suite are those simulations with a $\Sigma M_e$:$\Sigma M_p$ of 1:1 and an initial embryo mass of $0.025$ M$_\oplus$. In most figures and tables all suites are presented and they are arranged in a grid such that initial conditions corresponding to older oligarchic growth scenarios (i.e. further evolved before interruption by Jupiter's migration) are towards the right (larger initial embryo masses) and upwards (increasing embryo-planetesimal mass ratios). 

The inward migration of Jupiter interrupts the oligarchic growth phase of the protoplanetary disk. If the oligarchic growth phase were to reach completion, the individual planetary embryo mass would be proportional to $a^{3 (2-\alpha)/ 2}$, where $\alpha$ is the power-law index of the radial surface density $\Sigma$ profile of the solids in the disk: $\Sigma = \Sigma_0 r^{-\alpha}$, where $r$ is the distance from the Sun~\citep{Kokubo:2002dz}. While we use a surface density profile $\alpha = 3/2$, which is consistent with an estimate for the minimum mass solar nebula~\citep{Weidenschilling:1977kq,Hayashi:1981wz}, to determine the mass-orbit distribution of the embryos and planetesimals, the planetary embryos are not expected to reach their isolation masses. The timing of this interruption determines to first order the relative amount of mass in the embryo and planetesimal populations and the mass of the embryos. Since interior embryos grow faster than exterior ones, the final embryo mass distribution should be steeper than the final isolation mass distribution. For simplicity, we assumed that when Jupiter interrupts the oligarchic growth phase the embryo masses are identical. This is an acceptable approximation since the inward migration of Jupiter compresses the inner disk, displacing the embryos from their original runaway growth locations and emplacing them stochastically throughout the truncated disk. Before the inward migration of Jupiter, the disk is stable since the embryos are placed at constant ($\sim$5--10) mutual Hill radii from each other according to the total amount of mass in the disk and the prescribed $\alpha = 3/2$ surface density profile.

In the simulations, we adjust the total mass of the disc (embryos plus planetesimals) interior of Jupiter relative to the value used in~\citet{Walsh:2011co}. The reason is that a significant amount of mass initially outside of 1 AU is transported inside of 1 AU by being captured into a mean motion resonance with Jupiter during the inward migration phase. This process increases the density of solid material inside of 1 AU by about a factor of 2 and thus explains the enhanced density relative to all previous models assumed by~\citet{Hansen:2009ke}. The shepherding of bodies during the inward migration phase, though, is more efficient for planetesimals than for embryos. So, in order to have 2 Earth masses of material $\pm 10$\% inside of 1 AU at the end of the Jovian migration process, we need to start with 4.3 Earth masses between 0.7 and 3.0 AU if the ratio between the total mass in embryos and planetesimals is 1:1. However, the mass has to be increased to 4.9 Earth masses for the 2:1 ratio, 5.3 Earth masses for the 4:1 ratio and 6.0 Earth masses for the 8:1 ratio. These values are found empirically, after a few trials. 

We chose the free variables of the individual embryo mass and the ratio of the total embryo mass to total planetesimal mass because these are the primary protoplanetary disk parameters that control dynamical friction. It's good to recognize that the total embryo to total planetesimal mass ratio only applies to the total disk not at each semi-major axis in the disk, however the compression of the protoplanetary disk due to the inward migration of Jupiter appears to remove all of the structure built into the disk due to changing timescales with semi-major axis. So in order to reduce the size of the parameter space to explore, we simplified the initial disk structure.

There are two populations of planetesimals in these Grand Tack simulations. A population of inner planetesimals with a mass of $3.8 \times 10^{-4}$ Earth masses spread out between 0.7 and 3 AU (Jupiter begins at 3.5 AU before migrating inward). There is a disk of planetesimals exterior of 6 AU with individual masses of $1.2 \times 10^{-4}$ Earth masses~\citep{OBrien:2014vp}, and they are scattered inward when Saturn migrates outward. The initial orbits of the exterior planetesimals are consistent with scattering from a large extended disk or belts between the ice giants~\citep{OBrien:2014vp}. The exterior disk contributes very little mass ($\sim$1--3\%, although a significant amount of water) to the terrestrial planets and the inner disk, so it does not effect the protoplanetary disk dynamics or growth history of the terrestrial planets. The interior planetesimals are spaced a constant mutual Hill radii apart with a density such that they match the total amount of mass in the disk and the prescribed $\alpha = 3/2$ described above.

The simulations have been performed using the code Symba~\citep{Duncan:1998gn} and some with a newer version implementing OpenMP parallelization, kindly provided to us by H. Levison and D. Minton. We ran a total of 118 simulations for 150 Myr of evolution. Each simulation began with approximately a hundred embryos and two thousand planetesimals. The exact numbers depend on the mass of the embryos and the relative mass in the two populations. These simulations combined with the 8 simulations from~\citet{Walsh:2011co} and 16 from~\citet{OBrien:2014vp} make up the 142 simulations divided into 12 suites of initial conditions, as explained above. The Grand Tack creates on average 4 planets in each run regardless of the initial conditions with the exception of the 8:1-0.08 simulations which have a median of 3 planets. These simulations produced 238 Earth-like planets, since we are considering Earth-like planets to have masses within a factor of two of the Earth and orbits between the current orbits of Mercury and Mars.

\plotgrid{am-newGT6.0-8to1-0.25}
             {am-newGT6.0-8to1-0.5}
             {am-newGT6.0-8to1-0.8}
             {am-newGT5.3-4to1-0.25}
             {am-newGT5.3-4to1-0.5}
             {am-newGT5.3-4to1-0.8}
             {am-newGT4.9-2to1-0.25}
             {am-newGT4.9-2to1-0.5}
             {am-newGT4.9-2to1-0.8}
             {am-newGT4.3-1to1-0.25}
             {am-newGT4.3-1to1-0.5}
             {am-newGT4.3-1to1-0.8}
             {Each sub-panel shows planets from simulations that started from a different suite of initial conditions. The sub-panels are arranged according to initial embryo mass $M_e$ and the initial ratio between the total masses embryos and  planetesimals ($\Sigma M_e$:$\Sigma M_p$). Within each sub-panel, the mass of every planet from each simulation suite is shown as a function of semi-major axis. Each black dot locates the planet at its semi-major axis, and the black line shows its perihelion-to-aphelion excursion. Red dots show the real masses of Mercury, Venus, Earth and Mars, and the red lines show the width of their radial excursions. The green boxes highlight the regions that correspond to Mercury, Earth and Mars analogs.}
             {fig:mass}
             
All of these planets are shown in Figure~\ref{fig:mass}, which shows the mass and semi-major axis of each planet as a dot with a horizontal line through it. The horizontal line gives an indication of the eccentricity of the planet's orbit, because it ranges from the perihelion to the aphelion distances. These mass and orbit distributions look similar to the distributions obtained by~\citet{Hansen:2009ke} and~\citet{Walsh:2011co}. In each suite, the mass distribution has a characteristic arc-like structure, producing large mass planets between 0.8 and 1 AU and less massive planets interior and exterior of that region. This general trend is robust to our explorations of the $\Sigma M_e$:$\Sigma M_p$ vs. $M_e$ parameter space. However, even though the final mass distributions are similar, the evolution histories of the systems do change with these parameters as shown further in this paper.  

In order to discuss the similarities and differences between each suite of simulations and to compare the results with observational constraints, we define a number of planet-analog categories: Mercury, Earth and Mars. We have not distinguished between Earth and Venus, since it is difficult to determine where the mass and/or orbit boundary should be. For each analog, we bound the semi-major axis space with the semi-major axes  of the next planet, so Mercury analogs are located interior of 0.723 AU, Earth analogs are between 0.385 and 1.524 AU, and Mars analogs are exterior of 1 AU. We also select the analogs with mass criteria requiring that each analog is within a factor of two of each planet's mass. Using these boundaries, the median number of Earth analogs is two per simulation, and the initial values of $\Sigma M_e$:$\Sigma M_p$ and $M_e$  do not significantly effect that number.

The number of Mars analogs in each simulated Solar System is strongly dependent on the initial embryo mass. This result is a consequence of the lower mass boundary (0.0535 M$_\oplus$) for our definition of a Mars analog. If we lowered the boundary to 0.025 M$_\oplus$ then there would be about 1 Mars analog per simulation regardless of initial conditions. However, we think that matching the mass of Mars is important so such a low lower bound is not justified. In the following analysis we will use the 0.0535 M$_\oplus$ boundary.

All the simulations presented above are quite unsuccessful at creating Mercury analogs because they start with embryos that are roughly half the size, the same size or larger than Mercury itself and all collisions are perfect mergers. A mass-orbit trend pointing towards the real Mercury is clearly produced in the simulations, but typically our innermost planet is too far out and too massive with respect to the real Mercury. The 2:1-0.05 and 4:1-0.05 suites were able to produce a single Mercury analog and each of these is almost twice the mass of Mercury. In most cases there are no Mercury analogs because all planets exceed the upper mass boundary. The aforementioned trend suggests that the inner boundary of our disc is initially too far out. Discs extended further in but with a declining surface density may reproduce Mercury more satisfactorily. Embryos masses that increase with heliocentric distance and the implementation of hit and run collisions may also improve the results. All these attempts will be the object of another paper. 

\section{Role of dynamical friction}
The mass-orbit distributions shown in Figure~\ref{fig:mass} look very similar between the simulation suites demonstrating the overall robustness of the Grand Tack scenario to different disc conditions. However, the varied parameters: the ratio of the total mass in the embryo population to the total mass in the planetesimal population and the mass of the individual embryos, do change the dynamical excitation and orbit concentration of the final planets. We use two orbital structure statistics to diagnose these trends: the angular momentum deficit $S_d$ and the concentration statistic $S_c$.

\plotgrid{scsd-newGT6.0-8to1-0.25}
             {scsd-newGT6.0-8to1-0.5}
             {scsd-newGT6.0-8to1-0.8}
             {scsd-newGT5.3-4to1-0.25}
             {scsd-newGT5.3-4to1-0.5}
             {scsd-newGT5.3-4to1-0.8}
             {scsd-newGT4.9-2to1-0.25}
             {scsd-newGT4.9-2to1-0.5}
             {scsd-newGT4.9-2to1-0.8}
             {scsd-newGT4.3-1to1-0.25}
             {scsd-newGT4.3-1to1-0.5}
             {scsd-newGT4.3-1to1-0.8}
             {The angular momentum deficit $S_d$ and concentration statistic $S_c$ for each simulated Solar System in that suite. The dashed lines are the angular momentum deficit $S_d = 0.0018$ and concentration statistic $S_c = 89.9$ for the terrestrial planets in the Solar System. The sub-panels are arranged the same as in Figure~\ref{fig:mass}.}
             {fig:scsd}  

$S_c$ reflects in a single number that the mass-orbit distribution of the terrestrial planets is peaked in the center (Venus-Earth) and falls off at either end, effectively quantifying the pattern observed in Figure~\ref{fig:mass}. It is defined~\citep{Chambers:2001kt}: $$ S_c = \max \left( \frac{\sum_j m_j }{\sum_j m_j \left( \log_{10} \left(a /a_j\right) \right)^2 } \right),$$ where the summation is over the planets in the simulated Solar System and $m_j$ and $a_j$ are the mass and semi-major axis of the $j$th planet. $S_c$ is then the maximum of this function as $a$ is varied. The values of $S_c$ for each system are shown in Figure~\ref{fig:scsd} and the Solar System $S_c = 89.9$ is marked with a vertical dashed line. In general, the concentration statistic of Grand Tack simulations matches that of the observed terrestrial planets. However, the concentration statistic trends towards lower numbers as the total mass ratio of embryos to planetesimals grows and/or the mass of the individual embryos decreases. 

The angular momentum deficit determines the dynamical excitation of a system of planets relative to that same system of planets on circular orbits~\citep{Laskar:1997vw}. While perturbations between the terrestrial planets can exchange angular momentum, the overall deficit is basically conserved in the current Solar System because there is little exchange possible with the giant planets. The angular momentum deficit is defined as: $$ S_d = \frac{\sum_j m_j \sqrt{a_j} \left( 1- \sqrt{1-e_j^2 } \cos i_j \right)}{\sum_j m_j \sqrt{a_j}},$$ where the summation is again over the planets and $e_j$ and $i_j$ are the eccentricity and inclination of the $j$th planet~\citep{Chambers:2001kt}. The values of $S_d$ for each system are shown in Figure~\ref{fig:scsd} and the Solar System $S_d = 0.0018$ is marked by a horizontal dashed line. As shown in Figure~\ref{fig:scsd}, the angular momentum deficit increases as the ratio of the total mass in the embryo population to the the planetesimal population increases. 
It also increases if the individual mass of the embryos is smaller.

Unlike the $S_c$ parameter, the angular momentum deficit of the terrestrial planets, however, was not conserved during the late orbital instability of the giant planets, which presumably placed said planets on their final (current) orbits $\sim 4$ Gy ago~\citep{Brasser:2009ce,Brasser:2013wy}. Thus, the results of our simulations do not necessarily have to match the current angular momentum deficit of the terrestrial planets. However, the simulations of~\citet{Brasser:2009ce},  show that $S_d$ is unlikely to decrease during the giant planet instability, while it can easily increase from 0 to the current value. Thus simulations of terrestrial planet accretion resulting in a value $S_d$ larger than the current one are probably unsuccessful, whereas all those showing a smaller $S_d$ are potentially successful.~\citet{Brasser:2013wy} showed that most likely the value of $S_d$ of the terrestrial planets prior to the giant planet instability event was between 20\% and 70\% of the current value. 

Both the observed trends of $S_c$ and $S_d$ with the disc conditions are explained on the basis of dynamical friction. Dynamical friction is the process of damping the eccentricities and inclinations of the larger bodies (embryos) due to gravitational interactions with a swarm of smaller bodies~\citep[planetesimals;][]{Wetherill:1989ev,Stewart:2000kq}. Obviously, the more massive the planetesimal population, the stronger this effect is. The dependence on the individual embryo's mass is less obvious. However, it can be shown that the effect is stronger on a more massive embryo than on a smaller embryo~\citep{Stewart:2000kq}. Moreover, a system of few massive embryos is less prone to self-stirring than a more crowded one of smaller embryos, even if it carries the same total mass. Thus the excitation of the final planets becomes larger moving from bottom to top or right to left in the matrix of sub-panels of Figure~\ref{fig:scsd}. Notice that in Figure~\ref{fig:scsd} the results are roughly equivalent moving from the sub-panel in the bottom left corner to the one in the upper right corner, i.e. trading $M_e$ for $\Sigma M_e$:$\Sigma M_p$ at roughly equal dynamical friction. Larger excitation leads to a larger $S_d$ but also requires a less radially confined system to achieve orbital stability, i.e. a smaller value of $S_c$. 

Given our lack of knowledge of the value of $S_d$ at the end of the process of terrestrial planet accretion, it is difficult to conclude from Figure~\ref{fig:scsd} which disc conditions are more successful than others, although we can probably exclude those characterized by the weakest dynamical friction, with $M_e=0.025M_\oplus$ and $\Sigma M_e$:$\Sigma M_p\ge 4$. More stringent constraints will come from the analysis of the accretion timescales, in the next section. 

\begin{figure}
\centering
\includegraphics[width=0.49\textwidth]{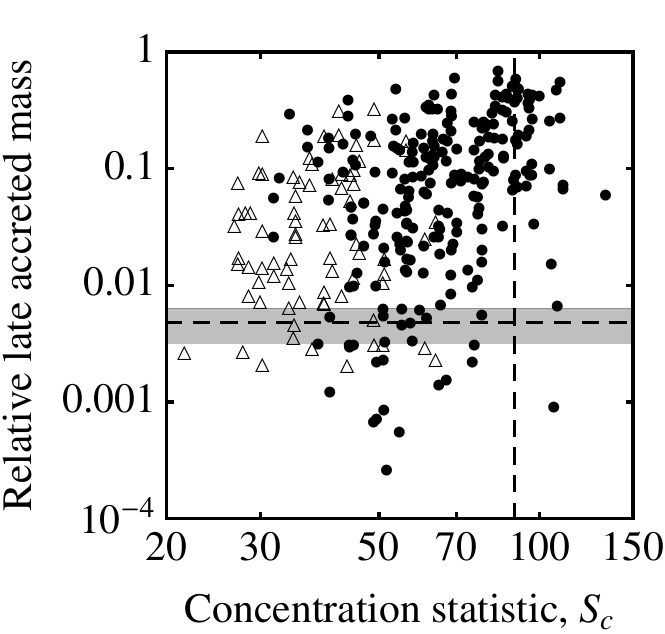} 
\includegraphics[width=0.49\textwidth]{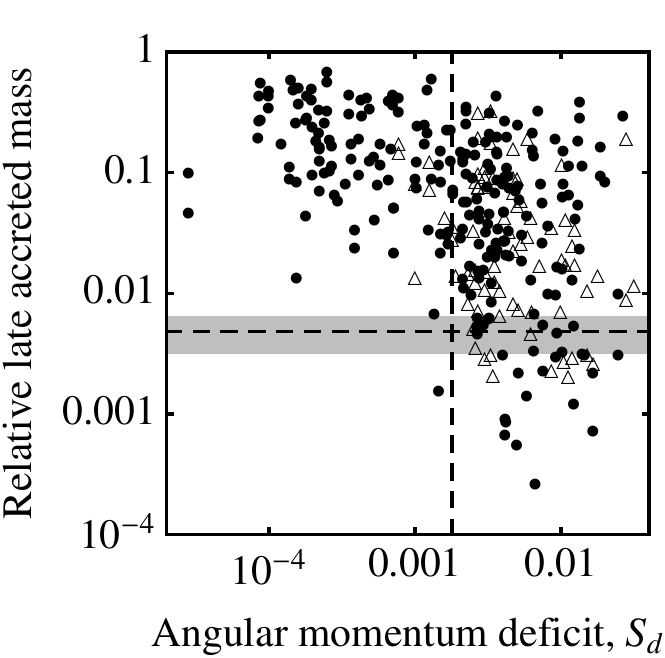}
\caption{\label{fig:scsdall} Both sub-panels show the relative late accreted mass for each Earth-like planet as a function of angular momentum deficit $S_d$ (left) and concentration statistic $S_c$ (right) for simulated Solar Systems. The points are from all of the Earth-like planets in the Grand Tack simulations and triangles are the values from published classical simulations~\citep{OBrien:2006jx,Raymond:2009is}. The vertical dashed lines are the angular momentum deficit $S_d = 0.0018$ and concentration statistic $S_c = 89.9$ for the terrestrial planets in the Solar System~\citet{Chambers:2001kt}. The horizontal dashed line is the estimated late accreted mass from the highly siderophile elements and the uncertainty is shown in gray: $M_{LA} = 1.8 \pm 0.6 \times 10^{-3}$ $M_\oplus$~\citet{Jacobson:2014cm}.}
\end{figure}

However, we can conclude that the Grand Tack simulations do a much better job in reproducing the real $S_d$ and $S_c$ values than the ``classic simulations''~\citep[e.g.][]{OBrien:2007ex,Raymond:2009is} that start from an extended disc of embryos and planetesimals and assume no migration of Jupiter's orbit (Figure~\ref{fig:scsdall}).  Indeed these simulations produce a value of $S_c$ way too small. This is obvious in view of the small-Mars problem that these simulations present, as discussed in the introduction. We also notice that the Grand Tack simulations can produce systems with an angular momentum deficit much smaller than the classical ones. 

It is also evident from Figure~\ref{fig:scsdall} that both the Grand Tack and the standard scenario can match the angular momentum deficit of the Solar System while simultaneously matching the late accreted mass, which is the planetesimal accretion after the last giant impact and estimated on the Earth from the highly siderophile elements~\citep{Jacobson:2014cm}. In both cases these are rarer outcomes, however we expect that the relatively large angular momentum deficit is an artifact of our simulation scheme and that in the future, more systems will match the constraints. In particular, future simulations will no longer need to assume perfect accretion, i.e. all collisions result in a single body, and instead collisions may generate debris as well as accretion or not lead to accretion at all~\citep[e.g. a hit-n-run collision;][]{Asphaug:2006gp}. Once debris is generated after giant impacts, this debris will act as newly generated planetesimals. Before those planetesimals are accreted by the planets or the Sun, they can damp the eccentricities and inclinations through a process of dynamical friction. The inclusion of this new process will shift the angular momentum deficits of the final systems to lower values, but early work has shown will not significantly effect the mass or semi-major axis distribution of the final planets~\citep{Kokubo:2010im,Chambers:2013cp}.

\section{Timing and other characteristics of the last giant (Moon forming) impact}
The dynamical friction within a planetary system controls the excitation of the final orbits by transferring momentum between embryos and planetesimals. The orbital excitation also determines in part the likelihood of giant impacts and the characteristics of those impacts (e.g. velocity). Higher eccentricities and inclinations correspond to increased random velocities and diminished mutual gravitational focusing. Consequently, decreased dynamical friction leads to a more protracted phase of embryo-embryo collisions and, thus, it increases the likelihood of having a later last giant (Moon forming) impact.

\plotgrid{LastGiantImpactsTimeMassPlot-newGT6.0-8to1-0.25}
             {LastGiantImpactsTimeMassPlot-newGT6.0-8to1-0.5}
             {LastGiantImpactsTimeMassPlot-newGT6.0-8to1-0.8}
             {LastGiantImpactsTimeMassPlot-newGT5.3-4to1-0.25}
             {LastGiantImpactsTimeMassPlot-newGT5.3-4to1-0.5}
             {LastGiantImpactsTimeMassPlot-newGT5.3-4to1-0.8}
             {LastGiantImpactsTimeMassPlot-newGT4.9-2to1-0.25}
             {LastGiantImpactsTimeMassPlot-newGT4.9-2to1-0.5}
             {LastGiantImpactsTimeMassPlot-newGT4.9-2to1-0.8}
             {LastGiantImpactsTimeMassPlot-newGT4.3-1to1-0.25}
             {LastGiantImpactsTimeMassPlot-newGT4.3-1to1-0.5}
             {LastGiantImpactsTimeMassPlot-newGT4.3-1to1-0.8}
             {The ratio of the impactor to total mass ratio for each last giant (Moon forming) impact on each Earth analog as a function of the time of that impact. The horizontal bands highlight the hypothesized lunar formation impactor to total mass ratios:equal sized colliders~\citep[orange,][]{Canup:2012cd}, standard and hit-n-run~\citep[magenta,][]{Canup:2008ff,Reufer:2012dz}, and rapidly rotating Earth~\citep[cyan,][]{Cuk:2012hj}. The vertical bands highlight predicted lunar formation times from geochemical evidence: early~\citep[red,][]{Yin:2002ex,Jacobsen:2005bj,Taylor:2009jr} and late~\citep[green,][]{Tera:1973tf,Carlson:1988jz,Touboul:2007is,Halliday:2008bo,Allegre:2008iq,Jacobson:2014cm}. The sub-panels are arranged the same as in Figure~\ref{fig:mass}.}
             {fig:timemass}  
             
The last giant impacts on each Earth analog in every simulation suite is shown in Figure~\ref{fig:timemass}. The times shown are measured from the first solids in the Solar System, but since the simulations began 0.6 Myr before the gas removal time, which we assumed occurred 3 My after the first solids~\citep{Haisch:2001bx}, we added 2.4 My to the time recorded in the simulations. As described previously, the grid of disc conditions is such that dynamical friction is progressively reduced from bottom to top and from right to left. Along both of these axes the influence of dynamical friction is clear; the time of the last giant impacts increases with reduced dynamical friction. There is considerable scatter because N-body dynamics are chaotic. 

\begin{figure}
\centering
\includegraphics[width=\textwidth]{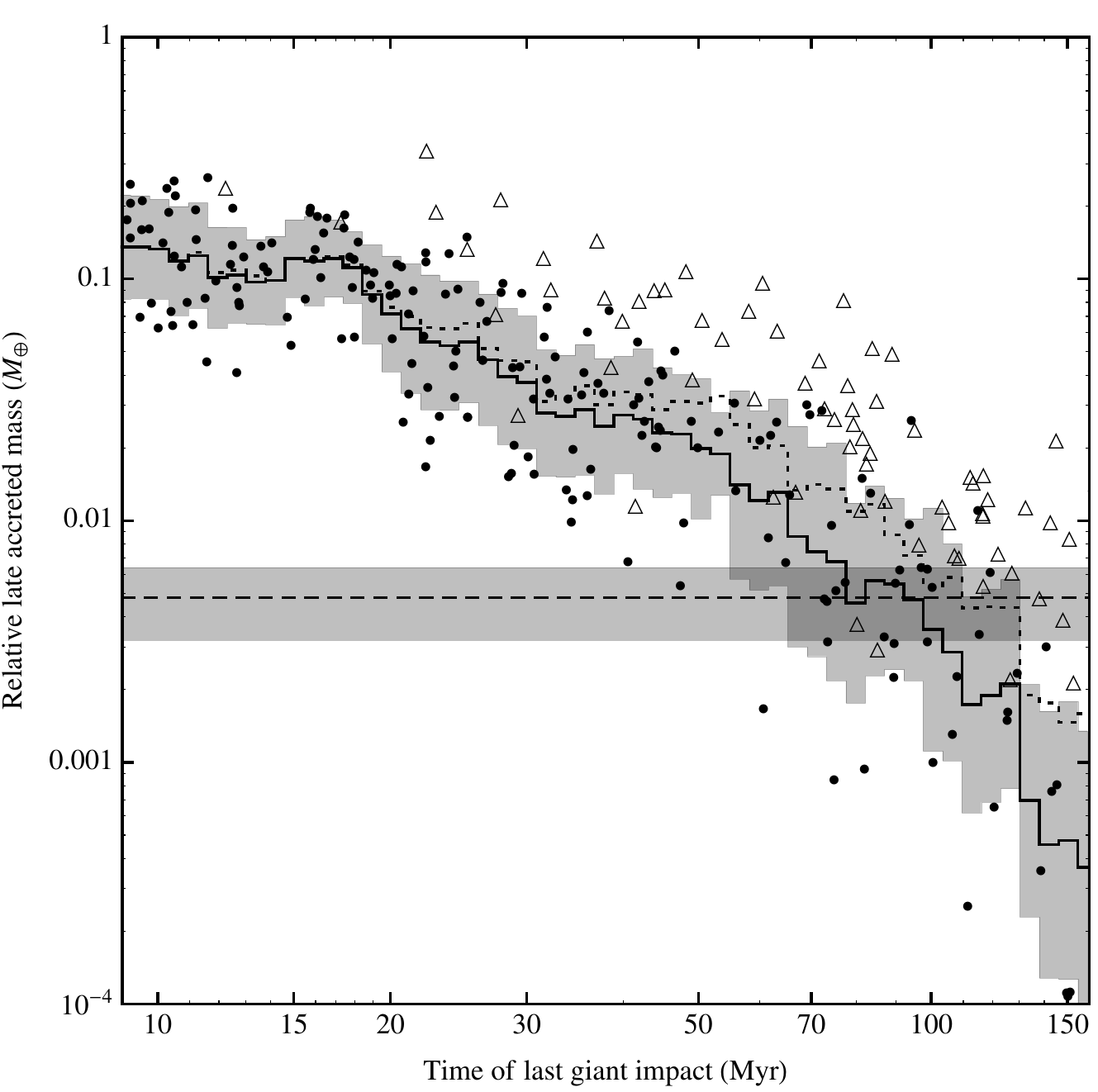} 
\caption{\label{fig:correlation} The late accreted mass (relative to the final mass of the planet) and the time of the last giant impact is shown for every Earth analog as a dot for Grand Tack simulations and a triangle for classical simulations~\citep{OBrien:2006jx,Raymond:2009is}. The downward arrow represents three Earth analogs that did not accrete any planetesimals after their final giant impact at nearly 150 My. Restricting ourselves solely to the Grand Tack simulations, the solid staircase is the running geometric mean over time of the late accreted mass and the gray region shows its 1-$\sigma$ uncertainty assuming that the recorded values are distributed log-normally. The dashed staircase is the same, but including also the results from classical simulations. The horizontal dashed line and gray region indicate the late accreted mass on the Earth $M_{LA} = 4.8 \times 10^{-2}$ M$_\oplus$ determined from the highly siderophile element abundance in the Earth's mantle.}
\end{figure}  

Despite this chaos, there is a strong correlation between the time of the last giant impact and the amount of late accreted mass, which is composed of the planetesimals accreted after the last giant impact~\citep{Jacobson:2014cm}. This correlation is shown in Figure~\ref{fig:correlation} and exists for both the Grand Tack and the classical simulations. Using an estimate of the late accreted mass $M_{LA} = 4.8 \times 10^{-2}$ M$_\oplus$ estimated from the ratio of the highly siderophile abundances of the Earth's mantle and chondritic meteorites~\citep{Chyba:1991cq,Walker:2009be}, the correlation dates the Moon forming impact to 95 $\pm$ 32 My~\citep{Jacobson:2014cm}. As extensively discussed in~\citet{Jacobson:2014cm}, the concentration of HSEs in the terrestrial mantle may not be a very accurate indicator of the actual amount late accreted mass. Nevertheless, it is unlikely that the late accreted mass on Earth exceeded 1\% of the Earth's mass, even if the most massive planetesimals were differentiated. This still provides a lower limit to the timing of the Moon forming event at 40 My at the 97.5\% confidence level.
             
Simulations of the lunar formation have resulted in a number of different predictions regarding the characteristics of the giant impact. The~\citet{Canup:2008ff} simulations have become the standard since they match dynamical constraints (e.g. angular momentum, mass in the disc), but they must rely on either Earth-disc processes or nearly geochemically identical progenitors to match the striking isotopic similarities between the Earth and the Moon. More recently, a hit-n-run collision between bodies very similar to the~\citet{Canup:2008ff} simulations has been proposed~\citep{Reufer:2012dz}, typically leading to a hotter disc, mixing more terrestrial material with impactor material, but with a smaller total mass. Also, it has been shown~\citep{Cuk:2012hj}, that the three-body evection resonance between the Earth, Moon and Sun may siphon angular momentum out of the Earth-Moon system (however see~\citet{Wisdom:2013wp} for a criticism of this idea). If so, two new dramatically different models have been proposed that can better match the isotopic constraints: the first calls for a smaller impactor on a more rapidly rotating Earth~\citep{Cuk:2012hj}, the other hypothesizes two similar sized bodies impacting each other~\citep{Canup:2012cd}. The impactor to total mass ratio at the time of the last giant impact for each Earth analog is shown in Figure~\ref{fig:timemass} and the hypothesized mass ratios are depicted as colored horizontal bands. 

The impactor to total mass ratio naturally increases as the initial embryo mass increases since many of the impactors are stranded embryos, which have not grown through embryo-embryo mergers. More subtly, the impactor to total mass ratio also increases as the total mass ratio of embryos to planetesimals decreases. The correlation in Figure~\ref{fig:correlation} explains this relationship. Earth analogs with the earliest last giant impacts also grow the most afterwards due to planetesimal accretion. Therefore, they are also the smallest at the time of the last giant impact and so the impactor and target are more likely to be of comparable size. Last giant impacts occurring near 95 My are, with a single exception, consistent only with scenarios that feature a Mars-mass or smaller impactor: the standard, rapidly rotating Earth, and hit-n-run scenarios. The two equal sized collider scenario is typically consistent only with very early last giant impacts, usually occurring earlier than 30 My.

\plotgrid{LastGiantImpactsVelImpParamPlot-newGT6.0-8to1-0.25}
             {LastGiantImpactsVelImpParamPlot-newGT6.0-8to1-0.5}
             {LastGiantImpactsVelImpParamPlot-newGT6.0-8to1-0.8}
             {LastGiantImpactsVelImpParamPlot-newGT5.3-4to1-0.25}
             {LastGiantImpactsVelImpParamPlot-newGT5.3-4to1-0.5}
             {LastGiantImpactsVelImpParamPlot-newGT5.3-4to1-0.8}
             {LastGiantImpactsVelImpParamPlot-newGT4.9-2to1-0.25}
             {LastGiantImpactsVelImpParamPlot-newGT4.9-2to1-0.5}
             {LastGiantImpactsVelImpParamPlot-newGT4.9-2to1-0.8}
             {LastGiantImpactsVelImpParamPlot-newGT4.3-1to1-0.25}
             {LastGiantImpactsVelImpParamPlot-newGT4.3-1to1-0.5}
             {LastGiantImpactsVelImpParamPlot-newGT4.3-1to1-0.8}
             {Each dot shows the impact velocity scaled by the escape velocity of the total mass and the impact parameter $b$ of the last giant (Moon forming) impact on each Earth analog. The dots are colored according to the impactor to total mass ratio predictions in Figure~\ref{fig:timemass}: equal sized colliders~\citep[orange,][]{Canup:2012cd}, standard and hit-n-run~\citep[magenta,][]{Canup:2008ff,Reufer:2012dz}, rapidly rotating Earth~\citep[cyan,][]{Cuk:2012hj}, and between these regions (gray). The polygonal regions are colored according to the same scheme to indicate the impact characteristics of each hypothesis. The standard scenario is the lower magenta box while the hit-n-run scenario is the upper magenta box. The sub-panels are arranged the same as in Figure~\ref{fig:mass}.}
             {fig:velparam}  
             
The various lunar formation scenarios also make predictions regarding the impact parameter and velocity. These predictions are shown as colored regions in Figure~\ref{fig:velparam}, while each last giant impact on an Earth analog is colored according to the impactor to total mass ratio region in in which they fall in Figure~\ref{fig:timemass}. Impacts with mass ratio exterior of any prediction are gray. The role of dynamical friction is clearly evident as the impact velocities increase from bottom to top and left to right. While many last giant impacts have impactor to total mass ratios similar to the rapidly rotating Earth scenario, these impacts can be too low velocity and are often too off axis with impact parameters much more consistent with the standard scenario. Since these simulations assume perfect merging, many of the higher impact parameter and velocity impacts are not accurately modeled. Likely these would be hit-n-run impacts. Some would be like those hypothesized in~\citep{Reufer:2012dz}, but others would bounce off the Earth~\citep{Asphaug:2006gp}. Simulations including hit-n-run collisions suggest that these bounced projectiles would likely return and strike the Earth again possibly at an impact parameter and velocity more similar to a hypothesized scenario.

In general, last giant impacts occurring on Earth analogs in the Grand Tack scenario can resemble hypothesized lunar formation scenarios. N-body simulations will likely not be able to rule out specific scenarios on their own, although equal sized colliders seem particularly rare. Trends tying the disc conditions at the time of giant planet migration to the properties of the last giant impact are messy but real. The strong correlation between the late accreted mass and the time of the last giant impact dates lunar formation to $\sim$95 My and, thus, constrains a large total mass ratios between embryos and planetesimals together with massive individual embryos or a low total mass ratio together with small individual embryos (i.e. relatively limited dynamical friction conditions). Interestingly the orbit statistics in Figure~\ref{fig:scsd} suggest the same conditions.  

Thus, we have a sort of degeneracy between total mass ratio between embryos and planetesimals and individual mass of the embryos. To break this degeneracy, we need to consider an additional constraint, namely the accretion history of Mars.

\plotgrid{MarsTimeMassPlot-newGT6.0-8to1-0.25}
             {MarsTimeMassPlot-newGT6.0-8to1-0.5}
             {MarsTimeMassPlot-newGT6.0-8to1-0.8}
             {MarsTimeMassPlot-newGT5.3-4to1-0.25}
             {MarsTimeMassPlot-newGT5.3-4to1-0.5}
             {MarsTimeMassPlot-newGT5.3-4to1-0.8}
             {MarsTimeMassPlot-newGT4.9-2to1-0.25}
             {MarsTimeMassPlot-newGT4.9-2to1-0.5}
             {MarsTimeMassPlot-newGT4.9-2to1-0.8}
             {MarsTimeMassPlot-newGT4.3-1to1-0.25}
             {MarsTimeMassPlot-newGT4.3-1to1-0.5}
             {MarsTimeMassPlot-newGT4.3-1to1-0.8}
             {The scaled growth history is shown for each Mars analog. From Hf-W evidence,~\citet{Nimmo:2007bg} predicts with 90$\%$ confidence that Mars growth history should not pass should not pass through the dark region and with $63\%$ confidence that it should not pass through the light red region. Using the same evidence but assuming an exponential growth model,~\citet{Dauphas:2011ec} predict that Mars accretion history should evolve inside the green region.}
             {fig:mars}  
             
\section{Consequences for Mars and the initial embryo mass}
The Hf-W system provides evidence that Mars must have formed very quickly~\citep{Nimmo:2007bg,Dauphas:2011ec} corroborated by the Fe-Ni system~\citep{Tang:2014ul}. Figure~\ref{fig:mars} shows with dark and right red boxes the regions of the mass vs. time space that should have been avoided by the growth history of Mars, according to the analysis of~\citet{Nimmo:2007bg}. The green region bounds the acceptable Mars' growth histories according to \citet{Dauphas:2011ec} and assuming exponential growth. The figure also reports the growth histories for Mars analogs that we observed in the simulations.

Clearly,  if the initial embryo mass is too small and requires multiple giant impacts, the Mars analogs do not grow in the predicted regions and cross the forbidden regions. This is because, even though the Grand Tack begins with a very violent concentration of the mass in the disc, the giant impacts do not happen fast enough. However, if the initial embryos are nearly Mars mass at the beginning of the giant impact phase, then they only need to accrete planetesimals in order to grow to full size. In this case, the Mars analogs grow in the predicted regions. This suggests that Mars is effectively a stranded embryo representative of the bodies existing in the planetesimal disc at the time Jupiter interrupted their orderly oligarchic growth. 

This result breaks the degeneracy mentioned at the end of the last section, and allows us to conclude that the ``good'' parameter space for the protoplanetary disc at the time of giant planet migration is in the top right corner of the domain we have explored. In fact, a re-analysis of all figures in this paper shows that the sub-panels at the top-right corner show a substantial number of simulations consistent with all the constraints that we have analyzed.

\section{Conclusions}

The Grand Tack Scenario invokes a specific migration pattern of Jupiter and Saturn during the proto-planetary disc phase in order to truncate the disc of planetesimals and planetary embryos at about 1 AU~\citep{Walsh:2011co}. Its results concerning terrestrial planet formation are much better than those obtained in classical simulations starting with a disc of embryos and planetesimals extended up to the orbit of Jupiter and neglecting any migration motion of the latter. In particular the distribution of masses vs. orbital radius of the terrestrial planets is much better reproduced and the final systems are characterized by a smaller angular momentum deficit.

We have shown in this paper that the results of the Grand Tack scenario can vary considerably in terms of Earth accretion history and Mars accretion history if we change two fundamental parameters that characterized the disc at the time of giant planet migration: the total mass ratio between embryos and planetesimals and the mass of the individual embryos.
	
Moreover, we have discovered a correlation between the timing of the last giant impact on Earth and the mass subsequently accreted from planetesimals (late accretion). The concentration of highly siderophile elements in the Earth's mantle and other considerations detailed in~\citet{Jacobson:2014cm} suggest that late accretion on Earth delivered less than 1\% of an Earth mass, with a most likely value around 0.5\%. The application of the aforementioned correlation constrains the Moon forming event to have happened later than 40 My after the  formation of the first solids, with a preferred date around 95 My.

This result on the last giant (Moon forming) impact time combined with constraints on the growth rate of Mars gauges the disc conditions at the time of giant planet migration. It shows that embryos should have had masses similar to the current mass of Mars and that most of the disc mass was in the embryos' population. This corresponds to the upper right corner of the grid of disc conditions in Figures~\ref{fig:mass},~\ref{fig:scsd},~\ref{fig:timemass},~\ref{fig:velparam} and~\ref{fig:mars}. Under these disc conditions the terrestrial region of the Solar System is born dynamically warm since dynamical friction is reduced relative to other considered disc parameters. Last giant impact velocities are typically enhanced, but remain smaller than a factor of two of the combined escape velocity, consistent with modern impact scenarios. The typical relative masses of the impactor do not support the equal sized collision scenario of~\citet{Canup:2012cd} but are consistent with all other scenarios~\citet{Canup:2008ff},~\citet{Reufer:2012dz} and~\citet{Cuk:2012hj}. 

\appendix

\section{Dynamical Friction}

We further study the role of dynamical friction by examining the time evolution of the angular momentum deficit. According to~\citet{Laskar:1997vw} and a simplification introduced by~\citet{Chambers:2001kt}, the angular momentum deficit of a population of bodies is calculated:
\begin{equation}
E_\text{all}(t) = \sum_j^N m_j(t) \sqrt{a_j(t)} \left( 1- \sqrt{1-e_j(t)^2 } \cos i_j(t) \right) = E_\text{embryos}(t) + E_\text{planetesimals}(t)
\end{equation}
\begin{equation}
E_\text{embryos}(t) = \sum_j^{N_\text{embryos}} m_j(t) \sqrt{a_j(t)} \left( 1- \sqrt{1-e_j(t)^2 } \cos i_j(t) \right)
\end{equation}
\begin{equation}
E_\text{planetesimals}(t) = \sum_j^{N_\text{planetesimals}} m_j(t) \sqrt{a_j(t)} \left( 1- \sqrt{1-e_j(t)^2 } \cos i_j(t) \right)
\end{equation}
where the subscript indicates the type of body being summed and $N$ is the number of bodies of that type. To define a statistic useful for comparing simulated terrestrial systems to the Solar System,~\citet{Chambers:2001kt} introduced a useful circular orbit normalization:
\begin{equation}
C_\text{all}(t) = \sum_j^N m_j(t) \sqrt{a_j(t)}  = C_\text{embryos}(t) + C_\text{planetesimals}(t) 
\end{equation}
\begin{equation}
C_\text{embryos}(t) = \sum_j^{N_\text{embryos}} m_j(t) \sqrt{a_j(t)} \qquad \qquad C_\text{planetesimals}(t) = \sum_j^{N_\text{planetesimals}} m_j(t) \sqrt{a_j(t)}
\end{equation}
Earlier, we compared the angular momentum deficit of the planetary system, i.e. surviving embryos, by using the~\citet{Chambers:2001kt} statistic $S_d = E_\text{embryos}(t) / C_\text{embryos}(t)$ at $t = 150$ Myr. Using the above defined terms, we further study in detail the angular momentum in the terrestrial protoplanetary disk and particularly as it evolves in time. We study that variation in two different ways.

\plotgrid{AllAMD-newGT6.0-8to1-0.25}
             {AllAMD-newGT6.0-8to1-0.5}
             {AllAMD-newGT6.0-8to1-0.8}
             {AllAMD-newGT5.3-4to1-0.25}
             {AllAMD-newGT5.3-4to1-0.5}
             {AllAMD-newGT5.3-4to1-0.8}
             {AllAMD-newGT4.9-2to1-0.25}
             {AllAMD-newGT4.9-2to1-0.5}
             {AllAMD-newGT4.9-2to1-0.8}
             {AllAMD-newGT4.3-1to1-0.25}
             {AllAMD-newGT4.3-1to1-0.5}
             {AllAMD-newGT4.3-1to1-0.8}
             {The normalized angular momentum deficit of the entire population of protoplanetary disk objects (green curves) and the separate embryo population (black curves) and planetesimal population for each simulated solar system. The planetesimal population is shown normalized to the circular orbits of the entire population of disk objects (red curves) as are the embryo and entire population curves, naturally. However, the planetesimal population is also shown normalized only to the angular momentum deficit of the planetesimal population on circular orbits (blue curves). The normalized angular momentum deficit of the terrestrial planets of the Solar System $S_d = 0.0018$ is shown as a blue dashed line and blue zone representing a factor of 2 about that line is also included. The green region corresponds to a prediction made by~\citet{Brasser:2013wy} for the inner Solar System before the giant planet instability, i.e. {\it Nice} model. We only have angular momentum evolution data for the~\citet{Walsh:2011co} simulations and not for the~\citet{OBrien:2014vp} simulations, so there are fewer 1:1-0.025 and 1:1-0.05 simulations. The sub-panels are arranged the same as in Figure~\ref{fig:mass}.}
             {fig:AllAMD}
First, we examine the evolution of the normalized angular momentum deficit in both the embryo and planetesimal populations in an absolute sense. It's useful to recognize that since $C_\text{planetesimals}(t) \ll C_\text{embryos}(t)$, then $C_\text{all}(t) \sim C_\text{embryos}(t)$. This means that $S_d(t) =  E_\text{embryos}(t) / C_\text{embryos}(t) \sim E_\text{embryos}(t) / C_\text{all}(t)$ (black curves) and so for convenience we only plot this second quantity along with the total normalized angular momentum deficit $E_\text{all}(t) / C_\text{all}(t)$ (green curves) and the planetesimal fraction of the normalized angular momentum deficit $E_\text{planetesimals}(t) / C_\text{all}(t)$ (red curves) in Figure~\ref{fig:AllAMD}. We also show just the normalized angular momentum deficit of the planetesimal population $E_\text{planetesimals}(t) / C_\text{planetesimals}(t)$ (blue curve). The primary plot in Figure~\ref{fig:AllAMD} displays time logarithmically since most of the changes in angular momentum deficit occur early in the history of the protoplanetary disk. To emphasize this, we show as an inset in Figure~\ref{fig:AllAMD} the identical data but with a linear time axis. It is clear that in most cases that $S_d(t)$ reaches a constant value after a few tens of millions of years.

In Figure~\ref{fig:AllAMD}, the total normalized angular momentum deficit (green curves) is naturally the sum of the embryo (black curves) and planetesimal (red curves) fractions of the angular momentum deficit. In every simulation suite, we see that for the first few hundred thousand years of the disk's history, the black and green curves overlap indicating that the embryo population dominates the angular momentum deficit. After this period, there is a transition until eventually the red and green curves overlap indicating that the planetesimal population dominates the angular momentum deficit population. 

\plotgrid{AllRelativeAMD-newGT6.0-8to1-0.25}
             {AllRelativeAMD-newGT6.0-8to1-0.5}
             {AllRelativeAMD-newGT6.0-8to1-0.8}
             {AllRelativeAMD-newGT5.3-4to1-0.25}
             {AllRelativeAMD-newGT5.3-4to1-0.5}
             {AllRelativeAMD-newGT5.3-4to1-0.8}
             {AllRelativeAMD-newGT4.9-2to1-0.25}
             {AllRelativeAMD-newGT4.9-2to1-0.5}
             {AllRelativeAMD-newGT4.9-2to1-0.8}
             {AllRelativeAMD-newGT4.3-1to1-0.25}
             {AllRelativeAMD-newGT4.3-1to1-0.5}
             {AllRelativeAMD-newGT4.3-1to1-0.8}
             {The relative angular momentum deficit of the embryo to the planetesimal populations for each simulated solar system. We only have angular momentum evolution data for the~\citet{Walsh:2011co} simulations and not for the~\citet{OBrien:2014vp} simulations, so there are fewer 1:1-0.025 and 1:1-0.05 simulations. The primary plot and the inset share identical data, however the time is shown logarithmically in the primary plot but linearly in the inset. The sub-panels are arranged the same as in Figure~\ref{fig:mass}.}
             {fig:AllRelativeAMD}
We examine this change from embryo to planetesimal dominance of the angular momentum deficit in detail with our second analysis technique---directly comparing the relative dynamical excitation in the system between the embryos and the planetesimals $R(t) = E_\text{embryos}(t) / E_\text{planetesimals}(t)$. Naturally, $R = 1$ means that the  angular momentum deficit compared to all bodies occupying circular orbits is equally distributed between the two populations.

In Figure~\ref{fig:AllRelativeAMD}, we show this ratio during the evolution of the protoplanetary disk. Now, we clearly see how angular momentum is transferred and lost between the two populations due to gas drag, dynamical friction and mass loss creating three distinct eras in the protoplanetary disk history. 

Since we begin the simulation with both the planetesimal and embryo populations on nearly circular orbits consistent with accretion in a gas disk, the ratio of angular momentum deficits is from 1 to 8 since the ratio of mass in the embryo to planetesimal populations $\Sigma M_e$:$\Sigma M_p$ also goes from 1 to 8. Then during the first 0.1 Myr of the simulation, Jupiter plows into the inner Solar System exciting everything, but due to the much larger effect of gas drag on planetesimals, only the embryos preserve their very excited orbits and the ratio of angular momentum deficit skyrockets. The effect of tidal gas damping on the embryos is included, but since this effect scales with mass, it is much less effective than the direct gas drag on the planetesimals. In summary, regardless of the initial value of $R(t)$, i.e. regardless of the initial $\Sigma M_e$:$\Sigma M_p$, the embryo population increases in excitation relative to the planetesimals by nearly an order of magnitude.

During the second era of angular momentum deficit evolution in the disk, as Jupiter and Saturn move outward and the gas begins to dissipate, dynamical friction circularizes the embryo orbits transferring that angular momentum to the planetesimal population. In a minority of cases, there appears a sudden plummet in the ratio in Figure~\ref{fig:AllRelativeAMD}. This is due to the rapid growth of the embryos in those disks, a process reminiscent of runaway growth. The largest embryo interacts with the most planetesimals, which circularizes its orbit the most enhancing its gravitational cross-section and acting as a feedback on its growth. Quickly, these rapidly growing embryos settle on nearly circular orbits while greatly exciting the planetesimals around them. Hence, the plummet in the ratio of angular momentum deficits. In the majority of cases, this process is more gradual and takes up to $\sim$50 Myr.

After $\sim$50 Myr, the number of terrestrial embryos has decreased to a nearly stable configuration and often only one or two more embryos will be lost or accreted in the system. From the black curves in the linear inset in Figure~\ref{fig:AllAMD}, we see that often these embryos do not represent a significant proportion of the angular momentum in the disk. This is primarily because these individual embryos are small and didn't participate in the runaway growth-like process. During this third era of the angular momentum deficit evolution of the disk, the angular momentum of the embryo population does not change much. As these embryos reach their final sizes, we transition to calling them planets and we are left with disks that at least according to angular momentum deficit can look very similar to the terrestrial planets of the Solar System.

Let's go back and focus on the planetesimal population. In Figure~\ref{fig:AllAMD}, the angular momentum deficit of the planetesimal population normalized to either itself (blue curves) or the entire disk (red curves) does a zig-zag. First, it rises during the inward migration of Jupiter due to a balance between excitation from Jupiter and gas drag, then it falls during the first part of the outward migration due to gas drag, and finally it rises again once the gas can no longer significantly damp the motion of the planetesimals. Both of the normalized angular momentum deficit curves (blue and red) for the planetesimal population rise until $\sim$20 Myr. Afterwards, the normalized angular moment deficit of the planetesimal population increases if its normalized to the angular momentum deficit of the planetesimal population on circular orbits (blue curves) but decreases if its normalized to the angular momentum deficit of the entire population on circular orbits (red curves). This divergence is due to a change in the rate the planetesimal population is being depleted.

\plotgrid{AllMass-newGT6.0-8to1-0.25}
             {AllMass-newGT6.0-8to1-0.5}
             {AllMass-newGT6.0-8to1-0.8}
             {AllMass-newGT5.3-4to1-0.25}
             {AllMass-newGT5.3-4to1-0.5}
             {AllMass-newGT5.3-4to1-0.8}
             {AllMass-newGT4.9-2to1-0.25}
             {AllMass-newGT4.9-2to1-0.5}
             {AllMass-newGT4.9-2to1-0.8}
             {AllMass-newGT4.3-1to1-0.25}
             {AllMass-newGT4.3-1to1-0.5}
             {AllMass-newGT4.3-1to1-0.8}
             {The mass in the terrestrial disk (black curves), embryo population (blue curves) and planetesimal (red curves) population. The sub-panels are arranged the same as in Figure~\ref{fig:mass}.}
             {fig:AllMass}
The mass in the terrestrial disks are shown in Figure~\ref{fig:AllMass} as well as the mass in the embryo and planetesimal sub-populations. First, notice that the total mass and embryo mass follow a very similar pattern regardless of initial conditions. About half the mass is lost during the inward migration of Jupiter, and then the embryo mass slowly asymptotes to the final mass as the planetesimal population is first mostly accreted between 0.1 and 20 Myr (the total mass does not change much) and then mostly scattered into the Sun after 20 Myr (the total mass decreases to meet the embryo mass).

Second, let's examine the region around $\sim$20 Myr, which we were discussing above. At this time, the mass loss from the planetesimal population becomes much more efficient. What has changed in this planetesimal population? The orbits of the planetesimals were being steadily excited by the embryo population through dynamical friction. Since the planetesimals started on nearly circular orbits, they could be excited for $\sim$20 Myr without dire consequences. However after $\sim$20 Myr, the orbits of the planetesimals start intersecting with the Sun in a significant numbers and the population rapidly depletes. Dynamical friction continues to heat the planetesimal population (blue curves in Figure~\ref{fig:AllAMD}) but since there is an ever decreasing amount of mass in the population, the planetesimal fraction of the total angular momentum deficit (red curves in Figure~\ref{fig:AllAMD}); this is why during the third era of the relative angular momentum deficit curves rise in Figure~\ref{fig:AllRelativeAMD}. Once the planetesimal population rapidly depletes, there is no longer the dynamical friction necessary to drive the runaway growth like process amongst the embryos.

\section*{Acknowledgment}
S.A.J. and A.M.. were supported by the European Research Council (ERC) Advanced Grant "ACCRETE" (contract number 290568).

\bibliography{biblio}
\bibliographystyle{apalike}

\end{document}